\documentclass[aps,preprint,showpacs,amsmath,amssymb]{revtex4}
\usepackage{graphicx,color}
\begin{document}
\title{The common origin of the family mass hierarchy and CP violation from flavour-dependent vacuum for quarks and leptons}
\author{Ying Zhang}
\email{hepzhy@mail.xjtu.edu.cn}
\affiliation{School of Physics, Xi'an Jiaotong University, Xi'an, 710049, China}
\date{\today}
\begin{abstract}
We rewrite the Yukawa interactions of the standard model in terms of flavour-dependent vacuum structure to address the family mass hierarchy and CP violation in both the quark sector and lepton sector. It is realized for the first time that the Lagrangian only includes the same number of degrees of freedom as phenomenological observables with no requirement of  extra particles or any new symmetry. The quark and lepton mass hierarchy arises as a natural result of the close-to-flat vacuum in flavour space. The CP violation in CKM and PMNS is explained as a general quantum phase between weak gauge eigenstates and Yukawa interaction states. The mechanism is proven by reproduction of all current quark/lepton mass data and the CKM/PMNS flavour mixings.
\end{abstract}
\pacs{12.15.Ff, 14.60.Pq, 12.15.Hh}
\keywords{ family mass hierarchy; CP violation; Yukawa interaction}
\maketitle

\section{Introduction}\label{sec.intro}
It is believed that there is some unknown mystery behind the Yukawa interaction in the Standard Model (SM).
Complex Yukawa couplings as a flavour-dependent matrix determine the fermion family mass hierarchy and flavour mixings with non-vanishing CP violation in both the quark sector and lepton sector.
The issues have been addressed with various new physics schemes including discrete symmetry\cite{Bhattacharyya2012PRL,King2016RPP}, extra gauge groups
\cite{Dias2018PRD,Martinez2018PRD},
extra dimension\cite{Maru2019PTEP,Archer2012JHEP, Nima2000PRD}, and  their combination \cite{Chen2009PLB,1901.09552}. Especially in the lepton sector, the Dirac CP violation has often been treated as a bridge between SM neutrinos and new physics on the high energy scale \cite{Shimizu2017JHEP}.
Instead of introducing new particles or symmetries, we rewrite the Yukawa interactions based on a close-to-flat flavour vacuum. Some pioneering research has investigated flavour physics from the complex vacuum expectation value (VEV) \cite{Branco1990PLB}. Differing from the space-time structure of vacuum, the flavour structure of vacuum in the letter means non-universal vacuum in flavour interactions.
The mechanism is inspired by the mathematic result that a flat  matrix (all elements equal to 1)  overlapping non-diagonal symmetric perturbation $\delta$ can yield hierarchical eigenvalues
\begin{eqnarray}
	\left(\begin{array}{cc}1 & 1+\delta \\ 1+\delta & 1\end{array}\right)\xrightarrow{diag}\left(\begin{array}{cc}d_1& \\ & d_2\end{array}\right),~
	\frac{d_1}{d_2}=\frac{\delta}{2}+\mathcal{O}(\delta^2).
\end{eqnarray}
Based on the results, we propose a assumption that the physical vacuum is nearly flat in flavour space.
With the difference from broken democratic-type mass matrices in \cite{Fritzsch1990PLB,Fritzsch2017CPC}, we only require real symmetric non-diagonal elements as perturbations.
The origin of CP violation can be considered separately from  a quantum phase between weak gauge eigenstates and Yukawa states. The rest of the paper is organized as follows:
in the next section, we show that a family mass hierarchy can arise naturally in terms of the mechanism in $3\times 3$ quark flavour space.
Then, to address CP violation, a Yukawa  phase is introduced in terms of the general quantum principle in sec.\ref{sec.CPV}.
The quantum phase between Yukawa interaction states and gauge eigenstates provides a common origin of CP violation. The rewritten Yukawa Lagrangian of quarks only includes 10 d.o.f., which correspond to 2 family total masses, 2 up-type quark hierarchies, 2 down-type quark hierarchies, 3 CKM mixings and 1 CP violation. The family-universal Yukawa coupling as a constant determines the total masses in the family. By reproducing all current quark masses and CKM at $3\sigma$, the validity is verified.
In sec.\ref{sec.PMNS},  the mechanism is generalized into lepton sector for normal hierarchy Dirac neutrinos, which is proved  to be successful in reproducing charged lepton masses, neutrino mass-squared difference $\Delta m^2_{ij}$ and the PMNS mixings. Finally, a summary is given in the last section.
\section{Quark mass hierarchy from perturbation vacuum}\label{sec.mass}
Assuming the physical vacuum has a close-to-flat structure in flavous space, its non-diagonal elements are parameterized by  3 real $\delta_{ij}^f$ as
\begin{eqnarray}({\bf v})^f_{ij}=v_0\left(\begin{array}{ccc}1 & 1+\delta^f_{12} &1+\delta^f_{13}\\ 1+\delta^f_{12} &1 & 1+\delta^f_{23} \\ 1+\delta^f_{13} & 1+\delta^f_{23} & 1\end{array}\right)
\label{eq.vij}
\end{eqnarray}
with SM VEV $v_0$, generation index $i,j=1,2,3$ and family index $f=d,u$ for down-type and up-type quarks.
The generation mechanism of non-diagonal vacuum perturbation is an open question. As a natural choice, multi-Higgs models can be realized the desired structure of the VEV by setting appropriate Higgs VEVs. Previous research has shown that  a democratic-type fermion mass matrix was generated from dimension five effective Lagrangian with two up-type Higgs doublets in \cite{FukuuraPRD2000,MiuraPRD2000, SogamiPTP1998}.
This kind of models also raises the question of FCNC in tree level.
Alternatively, close-to-flat vacuum may be provided by the SM itself.
Considering physical quantity renormalization, the Higgs potential is in fact a function of energy scale. As the minimum of potential, Higgs VEV is also affected by interaction scale.  Started from vanishing $\delta_{ij}^f$,  only the third generation is massive and the first two are massless,
which seed flavour violation. In the letter, ${\bf v}^f_{ij}$ is introduced as a theoretical assumption. We will pay more attention to the minimal parameterization of flavour structure.

\begin{figure}[htbp]
\begin{center}
\includegraphics[height=0.23 \textheight]{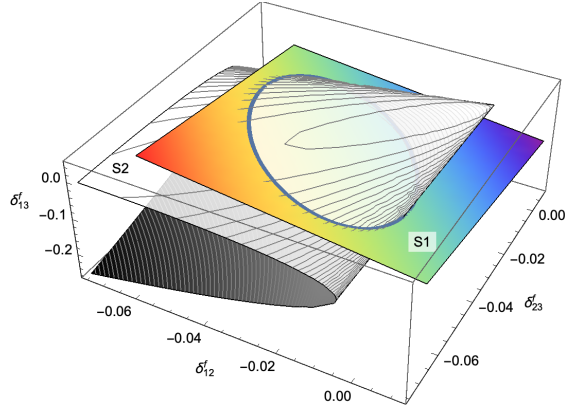}
\caption{{\bf Parameter space and family mass hierarchies.} Surfaces $S1$ (coloured rainbow) and $S2$ (coloured grey) correspond to eq.(\ref{eq.S}) and eq.(\ref{eq.P}), respectively. As an example,   $h_{12}^f$ and $h_{23}^f$ are valued at 0.05 and 0.02, respectively. The intersection line (coloured deep blue) is the parameter space allowed by the fermion mass hierarchies.}
\label{fig.explainVP}
\end{center}
\end{figure}

\begin{figure}[htbp]
\begin{center}
\includegraphics[height=0.23 \textheight]{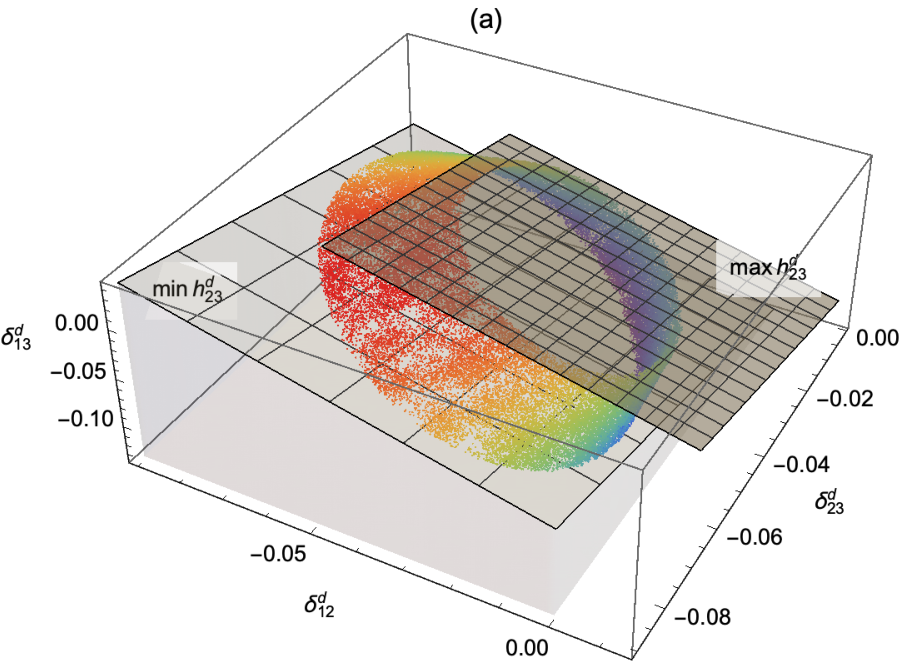}
\includegraphics[height=0.23 \textheight]{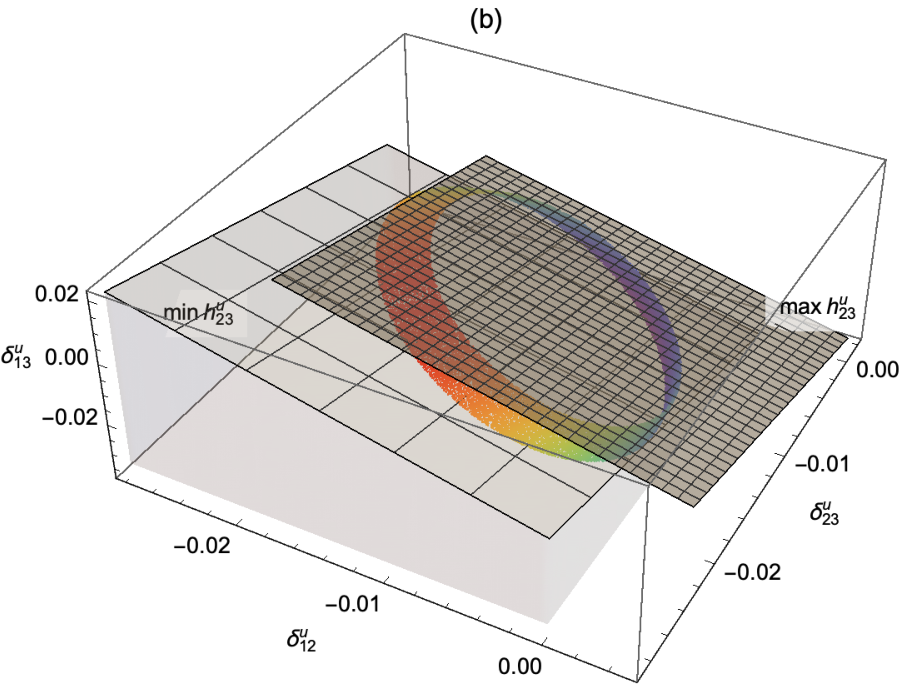}
\caption{Allowed vacuum perturbations  in the space of $(\delta_{12}^f,\delta_{23}^f,\delta_{13}^f)$ by quark mass data at $3\sigma$: (a) down-type quark and (b) up-type quark. The rainbow range is located between the maximum and minimum of hierarchy $h_{23}^f$.}
\label{fig.VPquark}
\end{center}
\end{figure}

Replacing the complex Yukawa coupling matrix $y^f_{ij}$ of the SM by a family-universal constant $y^f$, the quark mass terms become
\begin{eqnarray}
	-\mathcal{L}_m=\frac{y^d}{\sqrt{2}}\bar{d}_L {\bf v}^d_{ij} d_R+\frac{y^u}{\sqrt{2}}\bar{u}_L{\bf v}^u_{ij} u_R+H.c.
\label{eq.LagYukawa0}
\end{eqnarray}
From eqs.(\ref{eq.vij}) and (\ref{eq.LagYukawa0}), the quark mass matrix becomes
\begin{eqnarray}
	({\bf m}^f)_{ij}=\frac{y^fv_0}{\sqrt{2}}(1+\delta^f_{ij})
\end{eqnarray}
Physics masses $m_i^f$ can be obtained by diagonalizing matrix ${\bf m}^f$ by a real orthogonal rotation ${\bf W}_0^f$
\begin{eqnarray}
({\bf W}_0^f)^\dag {\bf m}^f{\bf W}_0^f={\rm diag}(m_1^f, m_2^f,m_3^f).
\label{eq.rotationw0}
\end{eqnarray}
At the zero-order approximation $\delta_{ij}^f=0$,  the eigenvalues of eq.(\ref{eq.rotationw0}) have one doubly degenerate zero eigenvalue and one non-zero $3y^fv_0/\sqrt{2}$. This case corresponds to flat vacuum in flavour space, in which only 3rd generation is massive and the others are massless. At the 1st-order approximation, the quark mass eigenvalues  are
\begin{eqnarray}
\frac{\sqrt{2}}{y^fv_0}m^f_{1,2}&=&\frac{1}{3}S\Big\{1\mp 2\sqrt{1-3P}\Big\}+\mathcal{O}(\delta^2)
\\
\frac{\sqrt{2}}{y^fv_0}m^f_3&=&3-\frac{2}{3}S+\frac{2}{27}S\Big(1-3P\Big)+\mathcal{O}(\delta^2)
\label{eq.masseigenvalues}
\end{eqnarray}
with parameters
\begin{eqnarray}
S&\equiv&-(\delta^f_{12}+\delta^f_{13}+\delta^f_{23})
\label{eq.S}\\
P&\equiv&\frac{\delta^f_{13}\delta^f_{23}+\delta^f_{12}\delta^f_{13}+\delta^f_{12}\delta^f_{23}}{(\delta^f_{12}+\delta^f_{13}+\delta^f_{23})^2}
\label{eq.P}
\end{eqnarray}
Defining the mass hierarchy $h_{ij}^f\equiv m_i^f/m_j^f$, $S$ and $P$  are determined by  $h_{12}^f$ and $h_{23}^f$ to leading order as
\begin{eqnarray}
S=\frac{9}{2}h_{23}^f+\mathcal{O}(h^2).
,~~~
P=\frac{1}{4}+\frac{h_{12}^f}{6}+\mathcal{O}(h^2).
\label{eq.hierarchy2SP}
\end{eqnarray}
Setting the family-universal Yukawa coupling $y^f=\frac{\sqrt{2}}{3v_0}\sum_{i} m_i^f$, it determines the total mass of the family.
In the space of $(\delta_{12}^f,\delta^f_{23},\delta_{13}^f)$, eq.(\ref{eq.S}) represents a plane $S1$ with intercept $-S$, which is dominated by $h_{23}^f$ in eq.(\ref{eq.hierarchy2SP}).
Eq.(\ref{eq.P}) represents a surface $S2$, which is dominated by hierarchy $h_{12}^f$.
The intersection is allowed by fermion mass hierarchies (see Fig.\ref{fig.explainVP} for details).
Through reproduction of all quark mass date reported by \cite{2018PDG} at $3\sigma$, the parameter space of vacuum perturbations is explored in Fig.\ref{fig.VPquark}.
Thus, it is shown that the close-to-flat flavour vacuum provides a natural mechanism to address the quark mass hierarchy.
\section{CP violation in CKM arising from the Yukawa phase}\label{sec.CPV}
In the transformation eq.(\ref{eq.rotationw0}), real vacuum perturbations  cannot provide  a complex phase of CP violation in CKM.
Now, we introduce a complex phase to the quark fields in the Yukawa interaction.
Due to the Yukawa interaction, not the gauge interaction, as a general case, a quantum phase between weak gauge eigenstates $q_{L,R}$  and Yukawa interaction states $q^{(Y)}_{L,R}$ is allowed,
\begin{eqnarray}
d_{L,R}^{(Y)}={\bf F}^d_{L,R} d_{L,R},~~~
u_{L,R}^{(Y)}={\bf F}^u_{L,R} u_{L,R}.
\label{eq.defineYukawaPhase}
\end{eqnarray}
The Yukawa phase has a diagonal form $({\bf F}_{L/R}^q)_{ij}=e^{i\alpha^q_{L/R,i}}\delta_{ij}$, which keeps the weak gauge symmetry invariant.
In Yukawa interaction states, the Lagrangian $\mathcal{L}_m$ is expressed in a simple form,
\begin{eqnarray}
	-\mathcal{L}_m=\frac{y^d}{\sqrt{2}}\bar{d}_L^{(Y)} {\bf v}^d_{ij}d_R^{(Y)}+\frac{y^u}{\sqrt{2}}\bar{u}_L^{(Y)} {\bf v}^u_{ij} u_R^{(Y)}+H.c.
\end{eqnarray}
The quark mass matrices including the Yukawa phase become
\begin{eqnarray}
{\bf m}^q=\frac{1}{\sqrt{2}}y^q{{\bf F}_{L}^q}^\dag {\bf v}^q{\bf F}_{R}^q,~~~q=d,u.
\label{eq.YukPhamass}
\end{eqnarray}
Obviously, unitarity Yukawa phases have no effect on quark mass eigenvalues.
However, they provide a quantum origin of the CP violation in quark flavour mixings. Making unitarity transformations	
$q_{L,R}=({\bf U}^q_{L,R})^\dag q^m_{L,R}$
to diagonalize ${\bf m}^q$, the quark weak current term becomes
	\begin{eqnarray*}
		-\mathcal{L}_W=\frac{g}{\sqrt{2}}\overline{u}^m_L\gamma^\mu{\bf U}_{CKM}d^m_L W_\mu^+
			+H.c.
	\end{eqnarray*}
where ${\bf U}_{CKM}\equiv {\bf U}^u_L{{\bf U}^d_L}^\dag$ is the CKM matrix.
In the mass eigenstates, a pure rephasing can still be performed
$q_{L,R}\rightarrow {\bf K}_q q_{L,R}$ to obtain the standard parameterization of $CKM$ \cite{Branco2012RMP}.
In terms of eqs.(\ref{eq.rotationw0}) and (\ref{eq.YukPhamass}),  by decomposing  ${\bf U}^q_{L,R}$   into real orthogonal rotation ${\bf W}_0^q$ and Yukawa phase ${\bf F}_{L,R}^q$ as
\begin{eqnarray}
{\bf U}_{L,R}^q=({\bf W}_0^q)^\dag {\bf F}_{L,R}^q,
\end{eqnarray}
the CKM matrix becomes
\begin{eqnarray}
{\bf U}_{CKM}=({\bf W}_0^u)^\dag {\bf F}_{L}^u({\bf F}_{L}^d)^\dag {\bf W}_0^d.
\label{eq.CKMdef}
\end{eqnarray}
Note that the CP violation arises due to the combination with the Yukawa phase  ${\bf F}_L^u({\bf F}_L^d)^\dag$. Because a total phase can be eliminated by rephasing, the Yukawa phases are generally parameterized by only 2  phases,
 \begin{eqnarray}
 {\bf {F}}^d_L=diag(1,1,1),~~~
{\bf {F}}^u_L=diag(1,e^{i\alpha^u_1},e^{i\alpha^u_2}).
 \end{eqnarray}
To reproduce the current CKM data, we choose a point $(\delta_{12}^d,\delta_{23}^d,\delta_{13}^d)$ from Fig. 2(a) to match another point  $(\delta_{12}^u,\delta_{23}^u,\delta_{13}^u)$ from Fig. 2(b) and then calculate $U_{CKM}$ by scanning all possible Yukawa phases $\alpha^u_1$ and $\alpha^u_2$. If the results agree with the CKM data at $3\sigma$, $(\delta_{12}^d,\delta_{23}^d,\delta_{13}^d)$ and   $(\delta_{12}^u,\delta_{23}^u,\delta_{13}^u)$ become a set of matched data. The matched results is shown in Tab.\ref{tab.ckmresult}. All quark masses, CKM mixings are  reproduced successfully in $1\sigma$.
Here the CKM mixing is chosen in the form of  the standard parametrization. The results have shown that the Yukawa phase between the Yukawa interaction state and weak eigenstates can be response for the origin of the CP violation in CKM.
\begin{table}[htp]
\caption{quark flavour parameters, mass eigenvalues and CKM mixings}
\begin{center}
\begin{tabular}{c|c}
\hline
\hline
flavour para. & $\left\{\begin{array}{l}
			y^u=\frac{\sqrt{2}}{3v_0}\sum_i m^u_i
			\\
			y^d=\frac{\sqrt{2}}{3v_0}\sum_i m^d_i
			\end{array}\right.\!\!,
	\left\{\begin{array}{l}
			\delta_{12}^u=-0.000059
			\\
			\delta_{23}^u=-0.016
			\\
			\delta_{13}^u=-0.0175
			\end{array}\right.\!\!,
	\left\{\begin{array}{l}
			\delta_{12}^d=-0.0055
			\\
			\delta_{23}^d=-0.065
			\\
			\delta_{13}^d=-0.0429
\end{array}\right.\!\!,
	\left\{\begin{array}{l}
		\alpha_1^u=-0.00091
		\\
		\alpha_2^u=0.090
		\end{array}\right.$
\\
\hline
matched results & $\left\{\begin{array}{l}
		m^u=1.705MeV
		\\
		m^c=1.293GeV
		\\
		m^t=173.0GeV
		\end{array}\right.,
	\left\{\begin{array}{l}
		m^d=4.331MeV
		\\
		m^s=103.3MeV
		\\
		m^b=4.172GeV
		\end{array}\right.,
	\left\{\begin{array}{l}
		s_{12}=0.226
		\\
		s_{23}=0.0428
		\\
		s_{13}=0.00360
		\\
		\delta_{CP}=68.1^\circ
		\end{array}\right.$
\\
\hline
exp. data \cite{2018PDG} & $\left\{\begin{array}{l}
		m^u=2.2^{+0.5}_{0.4}MeV
		\\
		m^c=1.275^{+0.025}_{-0.035}GeV
		\\
		m^t=173.0\pm0.4GeV
		\end{array}\right.\!\!,
	\left\{\begin{array}{l}
		m^d=4.7^{+0.5}_{0.3}MeV
		\\
		m^s=95^{+9}_{-3}MeV
		\\
		m^b=4.18^{+0.04}_{-0.03}GeV
		\end{array}\right.\!\!\!,\!
	\left\{\begin{array}{l}
		s_{12}=0.2244\pm0.0005
		\\
		s_{23}=0.0422\pm0.0008
		\\
		s_{13}=0.00394\pm0.00036
		\\
		\delta_{CP}=(73.5^{+4.2}_{-5.1})^\circ
		\end{array}\right.$
\\
\hline\hline
\end{tabular}
\end{center}
\label{tab.ckmresult}
\end{table}%
\section{Lepton PMNS mixings and Dirac CP violation}\label{sec.PMNS}
In the section, we generalize the flavour vacuum perturbation mechanism to the lepton sector with  3 massive Dirac neutrinos. After gauge symmetry breaking has taken place and lepton masses have been generated, the lepton weak current and mass terms can be written as
\begin{eqnarray}
	-\mathcal{L}&=&\frac{g}{\sqrt{2}}\bar{e}_L\gamma^\mu \nu_LW^-_\mu
	+\frac{y^e}{\sqrt{2}}\bar{e}^{(Y)}_L{\bf v}^e_{ij} e^{(Y)}_R
	+\frac{y^\nu}{\sqrt{2}}\bar{\nu}^{(Y)}_L{\bf v}^\nu_{ij} \nu^{(Y)}_R+H.c.
\end{eqnarray}
All formulas for quarks are conveniently generalized to the lepton sector.
Lepton masses have the same formula as eq.(\ref{eq.YukPhamass})
\begin{eqnarray}
    	{\bf m}_f=\frac{1}{\sqrt{2}}y^f({\bf F}_L^f)^\dag{\bf v}^f{\bf F}_R^f,~~~f=e,\nu
\end{eqnarray}
We have  similar results to those in the  quark sector:
\begin{itemize}
	\item[(1)] symmetric flavour vacuum  ${\bf v}^f$ is parameterized by 3 real non-diagonal perturbations;
	\item[(2)] in a family, two of the vacuum perturbations are determined by two mass hierarchies $h_{12}^f$ and $h_{23}^f$, and the third is free;
	\item[(3)] only 2 left-handed Yukawa phase differences have physical effects in PMNS;
	\item[(4)] the Yukawa phase provides the origin of the Dirac CP violation in PMNS;
	\item[(5)] including 2 family-universal Yukawa couplings $y^e$ and $y^\nu$, there are 10 parameters in lepton Lagrangian corresponding to 2 family total masses, 4 leptonic hierarchies and 4 PMNS observables.
\end{itemize}
Note that neutrino mass hierarchy $h_{ij}^\nu$ is determined by experimental data $\Delta m^2_{ij}$ after the lightest neutrino mass is initialized. Due to the high-accuracy charged lepton masses, the allowed vacuum perturbations are  distributed on a thin surface S1. Instead of scanning the parameter space of perturbations, we adopt a method of intersection line reconstruction.
Without loss of generality, we choose $\delta_{13}^{f}$ for $f=e,\nu$ as a free perturbation, and $\delta_{12}^f$ and $\delta_{23}^f$ are determined by charged lepton/neutrino mass hierarchy $h_{ij}^f$ in terms of eqs.(\ref{eq.S}), (\ref{eq.P}) and (\ref{eq.hierarchy2SP}), i.e., expressing $\delta_{12}^f$ and $\delta_{23}^f$ as  functions of $(h_{12}^f,h_{23}^f,\delta_{13}^f)$. The allowed vacuum perturbations lie on an intersecting circle of surface S1 and surface S2, like as one in Fig. \ref{fig.explainVP}.
The PMNS matrix is related to lepton unitarity transformations $l_{L,R}=({\bf U}_{L,R}^l)^\dag l^m_{L,R}$ by ${\bf U}_{PMNS}\equiv {\bf U}^e_L{{\bf U}_L^\nu}^\dag$.
Selecting a point $(\delta_{12}^{f},\delta_{23}^{f},\delta_{13}^{f})$ from  reconstructed data for $f=e$ and $f=\nu$, we can match them to current PMNS data \cite{2018PDG}  by running Yukawa phase $\alpha_i^\nu$ ($i=1,2$) from 0 to $2\pi$.
If there exists a Yukawa phase that makes the results lie in the $3\sigma$ range of PMNS parameters, then $(\delta_{13}^e,\delta_{13}^\nu)$ is recorded. A matched results is shown in Tab. \ref{tab.PMNSresult} for the case of $m_1^\nu=0.0001$eV.
\begin{table}[htp]
\caption{lepton flavour parameters, mass eigenvalues and PMNS mixings.}
\begin{center}
\begin{tabular}{c|c}
\hline
\hline
flavour para. & $\left\{\begin{array}{l}
			y^\nu=\frac{\sqrt{2}}{3v_0}\sum_i m^\nu_i
			\\
			y^e=\frac{\sqrt{2}}{3v_0}\sum_i m^e_i
			\end{array}\right.\!\!,
	\left\{\begin{array}{l}
			\delta_{12}^\nu=-0.150
			\\
			\delta_{23}^\nu=-0.441
			\\
			\delta_{13}^\nu=-0.095
			\end{array}\right.\!\!,
	\left\{\begin{array}{l}
			\delta_{12}^e=-0.01339
			\\
			\delta_{23}^e=-0.160
			\\
			\delta_{13}^e=-0.085
\end{array}\right.\!\!,
	\left\{\begin{array}{l}
		\alpha_1^\nu=-0.476
		\\
		\alpha_2^\nu=1.20
		\end{array}\right.$
\\
\hline
matched results & $\left\{\begin{array}{l}
		m^\nu_i=(0.0001,0.0086,0.0499)~eV
		\\
		\Delta m^2_{21}~[10^{-5}~eV^2]=7.40
		\\
		\Delta m^2_{31}~[10^{-3}~eV^2]=2.42
		\end{array}\right.\!\!,
	\left\{\begin{array}{l}
		m^e=0.51~MeV
		\\
		m^\mu=106.8~MeV
		\\
		m^\tau=1.775~GeV
		\end{array}\right.\!\!,
	\left\{\begin{array}{l}
		s_{12}^2=0.25
		\\
		s_{23}^2=0.38
		\\
		s_{13}^2=0.028
		\\
		\delta_{CP}=1.8\pi
		\end{array}\right.$
\\
\hline
exp. data  \cite{2018PDG}& $\begin{array}{c}
		3\sigma:
		\\
		\left\{\begin{array}{c}
		\Delta m^2_{21}~[10^{-5} eV^2]: 
        \\
        (6.93-7.96)
		\\
		\Delta m^2_{31}~[10^{-3}eV^2]: 
        \\
        (2.45-2.69)
		\end{array}\right.
		\end{array}\!\!\!\!,
	\left\{\begin{array}{l}
		m^e=0.5109989461(31)MeV
		\\
		m^\mu=105.6583745(24)MeV
		\\
		m^\tau=1776.86\pm0.12MeV
		\end{array}\right.\!\!\!\!,
    \begin{array}{c}
		3\sigma,~NH
		\\
		\left\{\begin{array}{l}
		s_{12}^2:0.250-0.354
		\\
		s_{23}^2:0.381-0.615
		\\
		s_{13}^2:0.0190-0.0240
		\\
		\delta_{CP}/\pi:1.0-1.9(2\sigma)
		\end{array}\right.
	\end{array}$
\\
\hline\hline
\end{tabular}
\end{center}
\label{tab.PMNSresult}
\end{table}%

\section{Conclusions}\label{sec.summary}
In conclusion, the family mass hierarchy is successfully explained in terms of flavour-dependent vacuum perturbation. In the mechanism, the CP violation in flavour mixings can arise as a quantum phase between weak gauge eigenstates and Yukawa interaction eigenstates.
The rewritten fermion mass terms greatly reduce the number of free parameters that appear in SM Yukawa couplings. There are $3+3$ vacuum perturbations (for up-type and down-type fermions) and 2 Yukawa phases in the quark/lepton sector, which generate $2+2$ mass hierarchies (for up-type and down-type fermions) of quarks/leptons and 3 mixings and 1 CP violation in CKM/PMNS. The Yukawa couplings as  family-universal constants determine the total fermion masses in the family. This is the first time that  the same number of d.o.f. in the description of the  flavour structure as that in physical observables has been realized without the help of new physics.
The validity of the mechanism is successfully  proven by reproduction of all fermion mass data and CKM/PMNS mixings in both the quark sector and lepton sector.
A remaining issue is to understand the dynamics of vacuum perturbation. As an open problem, some possible reasons may come from flavour-dependent interactions or the Higgs potential running with the interaction energy scale, which requires further investigation in the future.

\section*{Acknowledgements}
I wish to thank my collaborators Prof. Rong Li for discussions on the subject. This work is partially supported  by the Fundamental Research Funds for the Central Universities.


\begin{thebibliography}{10}
\bibitem{Bhattacharyya2012PRL}
	G. Bhattacharyya, I. de M. Varzielas, P. Leser,  Phys. Rev. Lett. 109(2012), 241604.
\bibitem{King2016RPP}
	S.F. King, C. Luhn, Rept. Prog. Phys. 76 (2013) 056201.
\bibitem{Dias2018PRD}
	A. G. Dias, J. Leite, D. D. Lopes, C. C. Nishi, Phys. Rev. D 98, 115017 (2018).
\bibitem{Martinez2018PRD}
	C.E. Diaz, S.F. Mantilla, R. Martinez, Phys. Rev. D 98(1), 015038 (2018).
\bibitem{Maru2019PTEP}
	N. Maru,  Y. Yatagai, Prog. Theor. Exp. Phys. 2019, 083B03(2019).
\bibitem{Archer2012JHEP}
	P. R. Archer, JHEP. 2012, 95 (2012).
\bibitem{Nima2000PRD}
	N. Arkani-Hamed, M. Schmaltz, Phys. Rev. D 61(2000) 033005.
\bibitem{Chen2009PLB}
	Mu-Chun Chen, K.T. Mahanthappa, Phys. Lett. B 681(2009) 444.
\bibitem{1901.09552}
	A.E.C. Hernández, S. Kovalenko, R. Pasechnik, I. Schmidt, Eur. Phys. J. C 79 (2019), 610.
\bibitem{Shimizu2017JHEP}
	Y. Shimizu, K. Takagia and M. Tanimotob, JHEP11 (2017) 201.
\bibitem{Branco1990PLB}
	G.C. Branco, J.I. Silva-Marocs, Phys. Lett. B 237 (1990) 3.
\bibitem{Fritzsch1990PLB}
	H. Fritzsch, J. Plankl, Phys. Lett. B 237(1990), 451.
\bibitem{Fritzsch2017CPC}
	H. Fritzsch, Zhi-zhong Xing, D. Zhang, Chin. Phys. C 41 (2017) 46.
\bibitem{SogamiPTP1998}
	I.S. Sogami,  K. Nishida,  H. Tanaka,  T. Shinohara, Prog. Theor. Phys. 99 (1998) 281.
\bibitem{FukuuraPRD2000}
	K. Fukuura, T. Miura, E. Takasugi, and M. Yoshimura, Phys. Rev. D 61(2000), 073002.
\bibitem{MiuraPRD2000}
	T. Miura, T. Shindou, E. Takasugi, and M. Yoshimura, Phys. Rev. D 63(2001), 053006
\bibitem{2018PDG}
	Particle Data Group (M. Tanabashi {\it et al.}), Phys. Rev. D 98 (2018) 030001.
\bibitem{Branco2012RMP}
	G.C. Branco, R.Gonzalez Felipe, F.R. Joaquim, Rev.Mod.Phys. 84 (2012) 515.
\end{thebibliography}
\end{document}